# From Static Bibliometrics to Dynamic Knowledge Graphs: An LLM-Powered Framework for Modernizing Science, Technology, and Innovation (STI) Analytics

MUHSEN HAMMOUD[1]

*1- Informatics Engineering Department, Informatics and Communication Engineering Faculty, Arab International University, Damascus, Syrian Arab Republic (m-hammoud@aiu.edu.sy)*

## Abstract

Bibliometric indicators — citation counts, h-indexes, co-authorship networks — have long anchored science, technology, and innovation (STI) analytics, yet suffer from temporal lag, semantic shallowness, and an inability to capture the non-linear dynamics of contemporary knowledge ecosystems. Dynamic knowledge graphs and large language models (LLMs) have each been proposed as remedies, but neither is sufficient alone: existing scholarly knowledge graphs remain largely static, while LLM-driven pipelines are prone to hallucination, opacity, and corpus bias without structured grounding.

This paper proposes a hybrid, symbolic-first framework integrating all three traditions under explicit methodological constraint. Organized across five layers — an open scholarly data backbone, a dynamic versioned knowledge graph, a constrained LLM-assisted semantic augmentation layer, a multi-layer validation pipeline, and an analytics layer — the framework positions LLMs strictly as generators of provisional candidate enrichments. Candidates become analytically admissible only after passing structural, evidentiary, comparative, and selective expert validation, with full provenance recorded at every stage. The analytics layer supports both established bibliometric indicators and extended graph-based analyses, including trend emergence detection, science-to-technology pathway mapping, and policy-oriented gap analysis.

The framework's central theoretical contribution is treating validation as the mediating principle between semantic flexibility and epistemic discipline, enabling STI analytics that is semantically richer and temporally more responsive than static bibliometrics while remaining aligned with the evidentiary standards of science-of-science research. Governance considerations addressing reproducibility, bias, and auditability are also discussed.

# 1. Introduction

Science, technology, and innovation (STI) systems have become increasingly complex and interconnected, characterized by tight coupling among scientific discovery, technological development, industrial application, and policy formulation (Schot & Steinmueller, 2018; OECD, 2024). In this landscape, effective analytics are essential for mapping emerging trends, identifying interdisciplinary synergies, allocating resources, and informing evidence-based decision-making. Traditional bibliometric approaches—relying on citation counts, impact factors, h-indexes, and co-authorship networks—have long served as the cornerstone of quantitative STI evaluation. Yet these methods, while quantifiable and scalable, suffer from inherent structural limitations that render them increasingly inadequate for capturing the dynamics of modern knowledge ecosystems.

Static bibliometrics primarily offer snapshot views of past performance, characterized by significant time lags between publication and citation accrual, which can delay the detection of emerging fields by years (Wang & Shapira, 2011). Citation-based metrics treat all links equally, ignoring semantic context, motivation, or quality of influence—leading to well-documented biases such as self-citation inflation, geographic/linguistic favoritism toward English-language and high-prestige journals, disciplinary skews, and overemphasis on quantity over substantive impact (Haustein & Larivière, 2014; Aksnes et al., 2019; Gøtzsche, 2022). Moreover, these approaches provide weak semantic resolution: they fail to distinguish between incremental refinements, paradigm shifts, negative citations, or interdisciplinary recombinations, resulting in superficial representations of knowledge flows (Smith, 1981). Even open and comprehensive alternatives like OpenAlex—a fully open scholarly knowledge graph indexing hundreds of millions of works, authors, institutions, and concepts—while dramatically improving accessibility and coverage over proprietary databases such as Web of Science or Scopus, still primarily deliver aggregated, largely static metadata unless actively extended with temporal and relational processing (Priem et al., 2022; Culbert et al., 2024; OpenAlex, 2026).

The growing complexity of STI systems exacerbates these shortcomings. Contemporary innovation often emerges from non-linear interactions across science–technology–policy–industry boundaries, involving rapid knowledge recombination, transient collaborations, and evolving conceptual trajectories (Schot & Steinmueller, 2018; Ba et al., 2025). Detecting such patterns requires representations that are not merely aggregative but relational, temporal, and semantically rich—qualities that static counts cannot deliver.

Dynamic knowledge graphs (KGs) offer a promising structural solution to these challenges. Unlike traditional citation networks, KGs model entities (e.g., concepts, authors, patents, funding sources) and their typed, directed relationships (e.g., "cites as method," "enables technology," "evolves into") within a flexible, queryable framework (Angioni et al., 2020). When extended with temporal dimensions—such as time-stamped edges, versioning, or decay functions—dynamic KGs enable tracking of knowledge evolution, emergence of communities, and foresight into innovation pathways (Weis, 2020; Aparicio et al., 2024). Open-access scholarly knowledge graphs like OpenAlex provide an excellent open foundation for such extensions, offering rich entity linkages and broad coverage that can be enriched dynamically to support STI analytics (Priem et al., 2022; Culbert et al., 2024; Haunschild & Bornmann, 2024). Recent applications demonstrate their utility in detecting emerging knowledge clusters and integrating heterogeneous STI data sources, moving analytics from retrospective description toward predictive and actionable insight (Bai et al., 2024).

Large Language Models (LLMs) represent a transformative development in this context, fundamentally changing the feasibility of constructing and querying such graphs at scale. LLMs excel at semantic extraction from unstructured scholarly text—identifying entities, inferring nuanced relations, synthesizing taxonomies, and reasoning over implicit connections that rule-based or traditional NLP methods struggle to capture (Agrawal et al., 2024; Li et al., 2025). Their ability to process vast corpora, generalize from few examples, and generate candidate triples or summaries dramatically lowers barriers to dynamic KG population, enrichment, and natural-language interfacing, enabling more timely and context-aware STI analytics.

Yet the promise of LLMs must be tempered by acknowledgment of their perils, particularly in high-stakes domains like science analytics where accuracy, transparency, and accountability are paramount. LLMs are prone to hallucinations—generating plausible but factually incorrect outputs—stemming from knowledge gaps, probabilistic decoding, and training data artifacts (Huang et al., 2025; Kazlaris et al., 2025). They can amplify biases present in their corpora (e.g., over-representing certain disciplines, geographies, or perspectives), propagate opacity through black-box reasoning, and introduce inconsistencies when applied to temporal or causal inference (Manda, 2025; Chandre et al., 2025). In scholarly KG construction, unchecked LLM use risks propagating erroneous triples, undermining trust in downstream analytics for policy or funding decisions (Li et al., 2025).

These limitations necessitate a principled middle path: a hybrid framework that harnesses LLMs' semantic strengths as targeted, task-specific tools within a controlled, symbolic architecture prioritizing validation, provenance, and oversight. By placing knowledge graphs at the core—providing structured grounding, consistency checks, and explainable paths—while restricting LLMs to suggestion generation, enrichment of ambiguous cases, or user-query translation (with mandatory post-validation via graph constraints, type rules, and confidence filtering), the framework mitigates risks while retaining benefits (Omar & Mohammed, 2025; Hussein et al., 2025).
This paper proposes such an approach. **Thesis**: A modern STI analytics framework must integrate the semantic understanding and generative capabilities of LLMs with the structural rigor, temporal expressiveness, and auditability of dynamic knowledge graphs, augmented by the validation mechanisms and methodological caution of traditional science-of-science studies. This hybrid path enables reliable modernization of STI analytics—from static, lagged metrics to dynamic, foresight-oriented representations—while safeguarding against the epistemic vulnerabilities introduced by probabilistic models.

## 2. Foundations and Related Work: Three Converging Literatures

Our framework sits at the intersection of three distinct but increasingly interconnected research streams: (i) the evolution—and persistent limitations—of STI analytics rooted in bibliometrics; (ii) the emergence of knowledge graphs as a structural paradigm for representing scientific knowledge; and (iii) the recent integration of large language models into scientific workflows. By critically examining each literature, we identify a clear gap: the absence of a principled, validated methodology that harnesses LLMs to **augment** rather than **replace** traditional STI analytics within a dynamic knowledge graph architecture emphasizing symbolic rigor, multi-layer validation, and temporal expressiveness. This section surveys each stream in turn, culminating in a synthesis that situates our contribution.

### 2.1. STI Analytics: From Static Indicators to Systemic Intelligence

The quantitative study of science and technology traces its origins to the mid-20th century, with foundational contributions from Price (1965) and Garfield (1972) establishing citation analysis as the dominant methodological paradigm. Subsequent decades witnessed a proliferation of indicators—the h-index (Hirsch, 2005), journal impact factors (Garfield, 2006), and co-authorship networks (Newman, 2001)—that collectively form what has been termed the "bibliometric toolkit" for research evaluation and policy (Moed, 2005; De Bellis, 2009).
Despite their widespread adoption, these indicators have been subject to sustained critique. The **Leiden Manifesto** (Hicks et al., 2015) and the **San Francisco Declaration on Research Assessment** (American Society for Cell Biology, 2012) crystallised concerns about the misuse of metrics, particularly their reduction of multidimensional research quality to single numbers, their vulnerability to gaming, and their inherent conservatism, which favours established paradigms over novel, interdisciplinary work. Bornmann and Mutz (2015) demonstrated that the exponential growth of scientific output exacerbates the temporal lag of citation-based indicators, rendering them ill-suited for real-time intelligence. Furthermore, Leydesdorff (1998) argued that citations encode a complex web of cognitive and social motivations that simple counts cannot capture—a limitation that subsequent work on citation content analysis has sought to address, but only at small scales (Small, 2011; Ding et al., 2014).
Parallel to these critiques, a systemic view of innovation has gained prominence. The **Triple Helix** model (Leydesdorff & Etzkowitz, 1998) and its extensions emphasise the recursive interactions among university, industry, and government. Coccia (2018) and Rotolo et al. (2015) highlight the role of general-purpose technologies and emerging technologies in reshaping innovation dynamics. Yet bibliometric methods remain largely focused on linear, dyadic relationships (e.g., citations, co-authorship), failing to capture the multi-actor, multi-relational nature of contemporary STI systems (Rafols & Meyer, 2010). This mismatch motivates the turn toward more expressive representational formalisms, among which knowledge graphs have emerged as the most promising.

### 2.2. Knowledge Graphs for Science: Structure without Semantics

Knowledge graphs (KGs)—graph-structured knowledge bases where entities are nodes and relationships are typed edges—have become a cornerstone of modern data integration and semantic search (Hogan et al., 2021). In the scientific domain, several large-scale KGs have been constructed to organise publication, patent, and researcher metadata. The **Microsoft Academic Graph** (Sinha et al., 2015), **OpenAlex** (Priem et al., 2022), **Aminer** (Tang et al., 2008), and the **Open Research Knowledge Graph** (Jaradeh et al., 2019) are prominent examples. These resources have enabled a new generation of scientometric analyses, from topic evolution tracking (Wang et al., 2020) to collaboration pattern mining (Jaradeh et al., 2019), with recent extensions like **CS-KG 2.0** (Dessì et al., 2025) scaling to 15 million papers and introducing temporal data for trend analysis.
However, existing scientific KGs are overwhelmingly **static** in both construction and content. Their schemas are typically fixed a priori, and their instances are populated through rule-based or dictionary-based extraction from structured metadata (e.g., author affiliations, journal names, citation links). While this ensures high precision, it imposes severe limits on semantic expressiveness. Relationships beyond the canonical few—cites, co-authorship, affiliation—are absent, and the rich conceptual content of scientific text (problems, methods, hypotheses, results) remains locked in unstructured natural language (Pan et al., 2024). Moreover, updates occur in periodic batches, introducing substantial latency and precluding the real-time or near-real-time analytics demanded by increasingly agile science policy and corporate R&D (Dessì et al., 2021).
Recent 2024–2026 advances have introduced **dynamic and temporal extensions** to scholarly KGs, including adaptive construction from unstructured text (ATOM: Lairgi et al., 2026, achieving high exhaustivity/stability via atomic fact splitting and dual-time modeling) and temporal reasoning for forecasting (e.g., TKG-LLM variants for time series-like prediction in scientific trends). These enable better foresight in STI contexts but still rely heavily on neural methods without strong symbolic validation, limiting auditability for policy use.
Efforts to enrich scientific KGs with deeper semantics have typically relied on either manual curation—unsustainable at scale—or supervised information extraction, which requires large, domain-specific annotated corpora that are costly to produce (Luan et al., 2018). Thus, while KGs provide an ideal **structural** container for STI intelligence, they currently lack the **semantic** engine needed to fill that container dynamically and at scale. This gap is precisely where large language models have recently entered the discourse.

## 2.3. Large Language Models in Scientific Analytics: Power and Peril

The arrival of transformer-based large language models (Vaswani et al., 2017; Devlin et al., 2019), culminating in frontier LLMs as of 2025–2026 (e.g., successors to GPT-4, Llama, and Qwen series), has fundamentally altered the landscape of natural language processing. LLMs exhibit remarkable capabilities in zero-shot and few-shot semantic understanding, text generation, and reasoning, leading to their rapid adoption across scientific domains (Bommasani et al., 2021). Within scientometrics and STI analytics, LLMs have been employed for tasks including literature summarisation (Koh et al., 2022), entity and relation extraction from scientific papers (Wadden et al., 2019; Luan et al., 2018), hypothesis generation (Hope et al., 2023; Kilicoglu et al., 2020), and content-based KG construction from citation networks (e.g., turning citation graphs "inside out" via LLM-derived taxonomies, 2026).
Despite these advances, a growing body of work documents the **perils** of deploying LLMs in high-stakes analytical contexts. **Hallucination**—the generation of plausible but factually incorrect or unsupported statements—is pervasive and difficult to eliminate (Ji et al., 2023; Zhang et al., 2025). LLMs also inherit and amplify biases present in their training corpora, potentially reinforcing existing inequalities in scientific visibility and funding (Bender et al., 2021; Lucy & Bamman, 2021). Furthermore, the opacity of their internal reasoning processes undermines trust and auditability—a fundamental requirement for policy-relevant analytics (Rudin, 2019; Shneiderman, 2020).
In response, several authors have proposed architectures that combine LLMs with external knowledge sources to ground generation in verifiable facts. **Retrieval-augmented generation** (RAG) (Lewis et al., 2020; Guu et al., 2020) is now a standard technique, and its application to scientific question answering is active (Ahrabian et al., 2023). Closer to our focus, recent work has explored using LLMs to **populate** knowledge graphs, either by generating triples directly (Ye et al., 2022) or by aligning textual mentions with KG entities (Wang et al., 2021). Multi-agent frameworks like **KARMA** (Lu et al., 2025) employ collaborative agents for entity discovery, relation extraction, schema alignment, and conflict resolution, achieving ~83% verified correctness on PubMed data while reducing conflicts via debate mechanisms. Neurosymbolic hybrids integrate LLMs with human-in-the-loop or symbolic validation in scholarly pipelines (e.g., Tsaneva et al., 2025 on CS-KG validation workflows, improving F1 by 5% with minimal human effort). Comprehensive surveys (e.g., Zhu et al., 2024; Bian et al., 2025 on LLM-empowered KG construction) highlight capabilities in reasoning and dynamic updates but note that most approaches treat the KG as passive and prioritize LLM dominance without systematic fusion with traditional bibliometric indicators or temporal versioning essential for STI analytics.

## 2.4. Synthesis and the Identified Gap

Synthesising these three literatures reveals a clear and consequential gap. Figure 1 visualizes this gap as the unoccupied intersection of three partially overlapping research streams — static bibliometrics, scholarly knowledge graphs, and LLM-driven approaches — and situates the proposed framework as the first principled attempt to occupy that space with symbolic rigor and bibliometric grounding. **Static bibliometrics** provide validated, auditable, but semantically shallow indicators with significant temporal lag. **Knowledge graphs** offer a powerful structural formalism for representing complex, multi-relational STI systems, but current scientific KGs are themselves static or emerging-dynamic without integrated symbolic safeguards. **Large language models** possess the semantic understanding and scalability needed to enrich KGs dynamically, but their outputs are unreliable without rigorous validation, and existing LLM-KG integration approaches (including multi-agent 2025–2026 systems) do not incorporate the safeguards nor the temporal perspective essential for credible STI analytics.
What is missing is a **principled, end-to-end framework** that:

- Preserves the verifiable foundation of traditional bibliometric indicators;
- Leverages LLMs **only** as constrained, task-specific semantic processors to generate candidate enrichments;
- Subjects those candidates to multi-layer validation (statistical, structural, provenance-based);
- Fuses validated enrichments into a dynamic, versioned knowledge graph;
- Enables novel STI analytics that are simultaneously scalable, semantically rich, and trustworthy.

This paper provides precisely such a framework—symbolic-first with targeted LLM augmentation under strict controls—differentiating from recent multi-agent (e.g., KARMA) or neurosymbolic hybrids by prioritizing bibliometric grounding and temporal auditability for policy-relevant foresight.

Table 1: Comparison of Core Literatures and Positioning of the Proposed Hybrid Framework

| Stream | Key Strengths | Persistent Limitations | How This Framework Addresses the Gap |
|---|---|---|---|
| Static Bibliometrics | Auditable, scalable indicators (h-index, citations) | Semantic shallowness, temporal lag, bias/gaming | Grounds dynamic KG in verified bibliometric signals |
| Scholarly KGs (e.g., OpenAlex, CS-KG 2.0) | Structured multi-relational representation | Static schemas, batch updates, limited semantics/temporal depth | Builds dynamic, versioned graph with temporal edges |
| LLM-Driven Approaches (e.g., frontier models) | Scalable semantic extraction, reasoning | Hallucinations, opacity, weak validation/temporal handling | Constrains LLMs to candidates + multi-layer symbolic validation |
| Emerging Hybrids (2024–2026, e.g., KARMA, Tsaneva et al.) | Reduced hallucinations via grounding/multi-agent | Often LLM-centric, limited STI/policy focus | Symbolic-first core with targeted augmentation + bibliometric fusion |

Table 1 and Figure 1 together highlight how our work bridges the streams responsibly: the table specifies capabilities and limitations at the level of individual contributions, while the figure shows the structural relationship between them and the gap that motivates this paper, avoiding over-reliance on probabilistic models while leveraging their strengths for modern STI analytics.

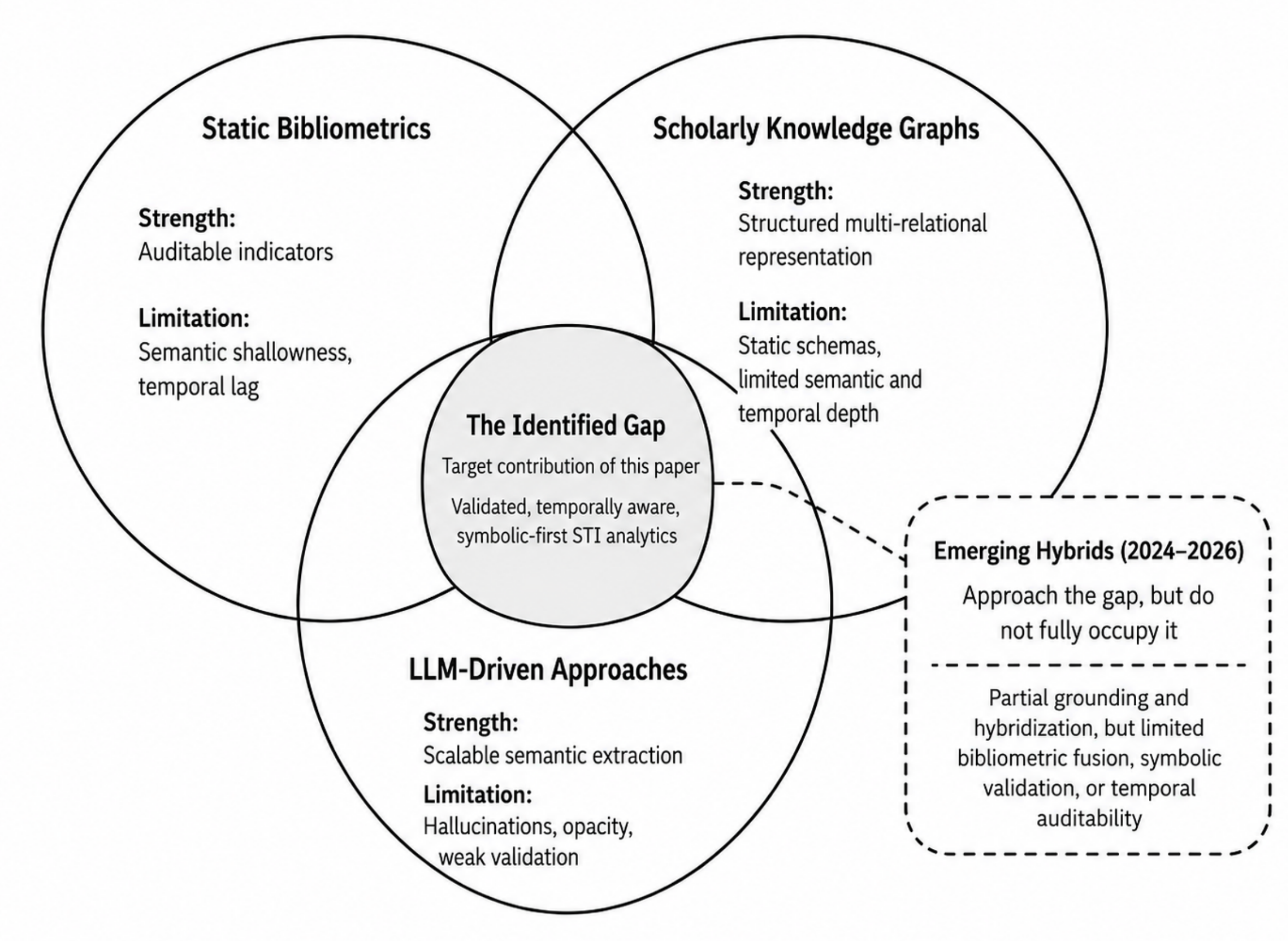

Figure 1 — The three converging literatures and the identified gap

# 3. Theoretical Foundations

The central premise of this paper is that contemporary STI analytics requires a representational and methodological shift. Traditional bibliometric approaches remain indispensable because they provide standardized, reproducible, and widely interpretable indicators of scientific activity. At the same time, their capacity to represent the semantic and temporal complexity of contemporary STI systems is limited (Rafols & Meyer, 2010; Glänzel & Thijs, 2012). The challenge, therefore, is not to replace established science-of-science methods, but to extend them in a way that preserves their evaluative discipline while improving their representational scope.

This section develops the theoretical basis for such an extension. It argues that a modern STI analytics framework should be understood as the integration of three complementary components: (i) validated indicator traditions from bibliometrics and scientometrics, (ii) knowledge graphs as relational and temporally extensible representational structures, and (iii) large language models as probabilistic tools for semantic augmentation under constraint (the process of enriching a base representation with additional, meaning-bearing relations extracted from unstructured sources). The central issue is how to combine these elements without sacrificing interpretability, reproducibility, or methodological caution.

## 3.1 STI analytics as a problem of representation and validation

STI analytics has historically depended on representations that reduce complex scientific processes into measurable units such as publications, citations, co-authorship ties, and institutional affiliations. These representations have enabled the construction of robust indicators for productivity, impact, collaboration, and interdisciplinarity. Their strength lies in standardization: they make large-scale comparison possible across fields, institutions, and periods.

However, the adequacy of a representation depends on the analytical questions being asked. Bibliometric representations are well suited to retrospective evaluation, but they are less effective when the task is to trace conceptual evolution, detect emerging combinations of ideas, or examine multi-actor interactions across science, technology, and innovation domains (Rafols & Meyer, 2010). In such contexts, the problem is not simply one of more data, but of representational mismatch. A publication-level or citation-level abstraction may be too coarse to capture the dynamics of methods, problems, claims, technologies, and institutional roles as interacting elements in an evolving system.

For this reason, the modernization of STI analytics can be understood first as a representational problem. Yet representation alone is insufficient. Since STI analytics often informs evaluative and policy-relevant decisions, any extension of representational capacity must also preserve validation standards. The theoretical problem, therefore, is dual: how to represent STI systems with greater semantic and temporal richness, and how to do so without weakening the evidentiary discipline that gives science-of-science methods their legitimacy.

### 3.2 Knowledge graphs as relational representations of STI systems

Knowledge graphs offer a natural response to the representational limitations of flat bibliographic structures. At a general level, a knowledge graph may be understood as a typed relational system in which heterogeneous entities are connected through explicitly defined relations. In the STI domain, these entities may include publications, authors, institutions, concepts, datasets, methods, patents, funding instruments, or policy documents. Their relations may include citation, authorship, affiliation, topical similarity, methodological reuse, sectoral linkage, or temporal succession.

The analytical significance of knowledge graphs lies in their ability to move beyond homogeneous ties. Unlike conventional citation networks, they allow different kinds of relationships to coexist in the same structure, thereby supporting richer interpretations of knowledge production and diffusion. They also provide a more suitable basis for integrating heterogeneous sources, which is especially important in STI settings where scientific, technological, and institutional processes interact.

A further advantage is temporal extensibility. A knowledge graph need not be treated as a static object; rather, it can be regarded as an evolving relational representation whose nodes, edges, and attributes change over time (Wu et al., 2024; Alam et al., 2024). This dynamic perspective is particularly relevant for STI analytics, where the emergence of concepts, reconfiguration of collaborations, and diffusion of technologies are inherently temporal phenomena. In this sense, dynamic knowledge graphs are not merely a technical refinement but an epistemically more appropriate representation for questions concerned with evolution, transition, and early signals.

At the same time, the usefulness of such graphs depends on the validity of the relations they contain. A graph may be structurally expressive yet analytically weak if its semantic content is noisy, inconsistent, or opaque. This places the problem of graph construction at the center of the framework.

### 3.3 Large language models as instruments of semantic augmentation

The recent emergence of large language models changes the practical feasibility of semantic enrichment in scholarly and STI data systems. These models can identify entities, propose relations, normalize terminology, summarize conceptual content, and support natural-language interaction with complex corpora. In this respect, they offer capabilities that are highly relevant to the longstanding bottleneck in scholarly knowledge graph construction: the conversion of unstructured text into structured, machine-actionable representations (Zhu et al., 2024; Bian et al., 2025).

However, from the standpoint of scientometrics, the significance of LLMs should be stated carefully. Their contribution is not that they provide a new source of validated indicators. Rather, they provide a means of generating semantic candidates at a scale that was previously difficult to achieve through manual curation or narrowly supervised extraction pipelines. Their role is therefore auxiliary and inferential, not authoritative.

This distinction is essential because LLM outputs are probabilistic rather than deterministic. They may be useful for proposing latent conceptual relations or extracting structured elements from text, but they do not in themselves satisfy the requirements of reliability, traceability, or reproducibility expected in evaluative analytics. Their outputs must be treated as provisional interpretations of source material rather than as established facts.

Accordingly, the theoretical place of LLMs in STI analytics is not at the level of final judgment or metric definition, but at the level of semantic augmentation. They can expand the range of relations available for analysis, but only if their outputs remain subordinate to explicit validation procedures and constrained by external structure.

### 3.4 Validation as the mediating principle

The key theoretical challenge is therefore not whether LLMs can enrich STI representations, but under what conditions such enrichment becomes analytically acceptable. The answer proposed here is validation. Validation functions as the mediating principle between semantic flexibility and epistemic discipline.

In the present context, validation should be understood as multi-layered. At a minimum, it includes structural validation, whereby candidate entities and relations are checked against schema constraints, type restrictions, and ontological consistency. It also includes evidentiary validation, whereby proposed enrichments are assessed against source texts, corroborating records, or repeated occurrence across documents. A further layer is comparative validation, in which graph-derived patterns are examined in relation to established bibliometric signals rather than treated as independent replacements for them. Finally, some categories of high-impact or uncertain enrichments may require expert review (Tsaneva et al., 2025; Lavrinovics et al., 2025).

This layered view is important because it clarifies that the framework does not seek to substitute probabilistic language generation for science-of-science methodology. Instead, it situates LLM-based enrichment within a controlled inferential pipeline. In this formulation, bibliometric indicators retain their role as validated baselines; knowledge graphs provide the representational substrate; and LLMs serve as constrained mechanisms for extending semantic coverage. Validation is what makes their combination methodologically coherent.

### 3.5 Versioning, temporality, and analytical accountability

A further requirement follows from the temporal nature of STI analytics: graph enrichment must be historically accountable. If the graph evolves over time, its states must remain reconstructable. This implies versioning, provenance tracking, and explicit association between graph updates and their evidentiary sources.

This requirement is not merely technical. In scientometric terms, reproducibility depends on the ability to determine which representation gave rise to which analytical result. If a trend, cluster, or policy-relevant inference is derived from a graph, that graph state must be identifiable, and its construction process must be inspectable. Otherwise, the transition from static indicators to dynamic representations risks weakening rather than strengthening analytical credibility.

Temporal accountability also addresses a specific weakness of purely generative systems: the instability of outputs across prompts, model versions, and contextual settings. By anchoring enrichments within versioned graph states, the framework restores a degree of reproducibility that would otherwise be difficult to maintain in LLM-assisted workflows.

### 3.6 Theoretical implication for modern STI analytics

Taken together, the preceding arguments suggest that the modernization of STI analytics should not be framed as a move away from bibliometrics, but as a layered extension of its analytical logic. Bibliometrics contributes standardized and validated measurement traditions. Knowledge graphs contribute relational and temporally extensible representations. LLMs contribute scalable semantic augmentation. None of these components is sufficient on its own.

The theoretical implication is that a credible next-generation STI analytics framework must combine representational richness with evaluative restraint. It must be able to incorporate semantically meaningful relations without abandoning the validation principles that underpin science-of-science inquiry. In this sense, the framework proposed in this paper is not an alternative to bibliometrics, but an attempt to place bibliometric reasoning within a broader representational environment suited to contemporary STI systems.

This theoretical position provides the basis for the framework presented in the next section. That framework operationalizes the distinction developed here between structural representation, semantic augmentation, and validation, and shows how they may be integrated in a dynamic knowledge graph architecture for STI analytics.

## 4. A Framework for LLM-Assisted Dynamic Knowledge Graph Analytics in STI

Building on the theoretical position developed above, this section presents the proposed framework for modernizing STI analytics. The framework is designed around a simple methodological principle: *semantic expansion should take place under structural and evidentiary constraint*. Its purpose is not to displace existing bibliometric practice, but to extend it by enabling richer representations of STI systems while preserving traceability, validation, and interpretability.

The framework consists of five analytically distinct layers: an open data backbone, a dynamic knowledge graph layer, an LLM-assisted semantic augmentation layer, a validation and governance layer, and an analytics layer. These layers are conceptually separable even if, in practice, they may partly overlap. Figure 2 provides an architectural overview of all five layers and their directional relationships. It shows that data flows upward from the open data backbone through the knowledge graph and augmentation layers to the analytics layer, while a critical feedback loop runs from the validation layer back to the augmentation layer, ensuring that only admissible enrichments propagate to the graph.

A secondary feedback path from the analytics layer to the knowledge graph layer represents versioned graph updates following each analysis cycle. The subsections below describe each layer in detail.

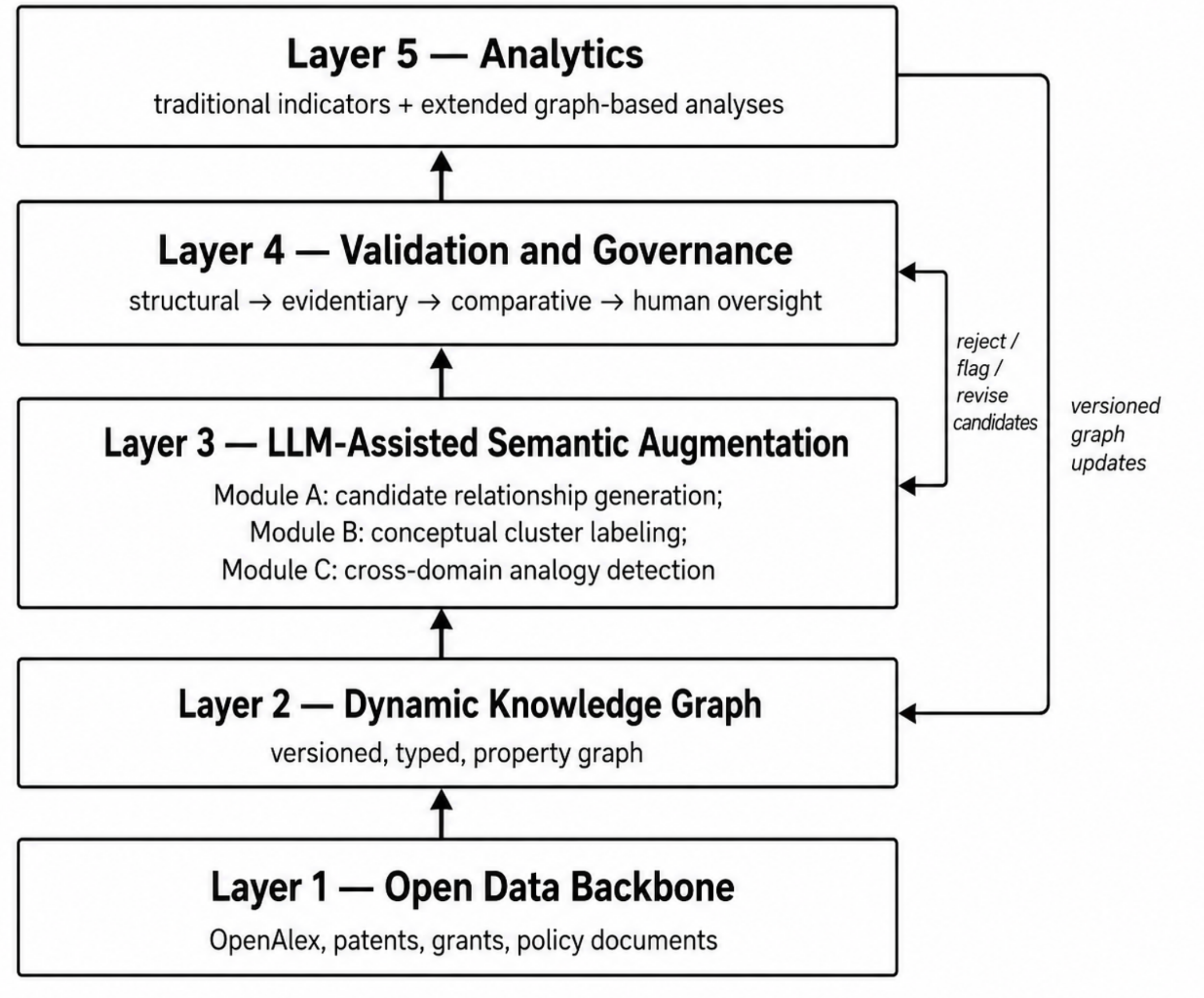

Figure 2 — The five-layer framework architecture

## 4.1 Open Data Backbone

The framework begins with an open, continuously updated scholarly data backbone. In practical terms, infrastructures such as OpenAlex provide the most suitable point of departure because they offer large-scale, interoperable, and reproducible representations of core scholarly entities and relations (Priem et al., 2022). At this level, the framework draws on established metadata such as publications, authors, institutions, venues, concepts, citations, and, where available, links to grants, patents, datasets, or policy-related documents.

This layer serves two purposes. First, it provides the initial structural substrate from which richer semantic relations may later be constructed. Second, it anchors the framework in a transparent and inspectable data environment. This is especially important for scientometric use, since any subsequent enrichment should remain connected to an identifiable source infrastructure rather than emerging from an opaque, self-contained workflow.

The framework therefore assumes that semantic enrichment should proceed from openly documented entities and relations rather than from unconstrained generation. In this sense, the open data backbone functions not only as a source of input data, but also as an initial condition for analytical accountability.

## 4.2 Dynamic Knowledge Graph Layer

On top of this backbone, the framework constructs a dynamic knowledge graph that represents STI systems as evolving, multi-relational structures. Unlike conventional bibliographic representations, this graph is not limited to publication-level metadata or citation ties. It is formally defined as a typed, property-graph structure with explicit provenance.

Let G = (V, E, τ, ρ, λ, θ) denote the dynamic knowledge graph, where:

- V is a set of entity nodes (e.g., publications, patents, authors, institutions, concepts, methods, problems).
- E ⊂eq V × Vis a set of directed edges.
- τ: V → T assigns a type to each node from a type ontology T.
- ρ: E → R assigns a relation type to each edge from a relation ontology R.
- λ: E → L attaches provenance metadata (source document, extraction method, timestamp, validation status).
- θ: E → ℝ optionally assigns a confidence score or weight.

Figure 3 instantiates this formal definition with a concrete STI example. It shows a small subgraph in which a publication node is connected to a method node via a USES_METHOD edge, to a concept node via an ADDRESSES_PROBLEM edge, and to a patent node via an ENABLES_APPLICATION edge. Each edge carries provenance metadata illustrating the λ component — source document identifier, extraction timestamp, and validation status — alongside a confidence score representing θ. A dashed boundary demarcates the snapshot state Gt, visualizing the versioned nature of the graph.

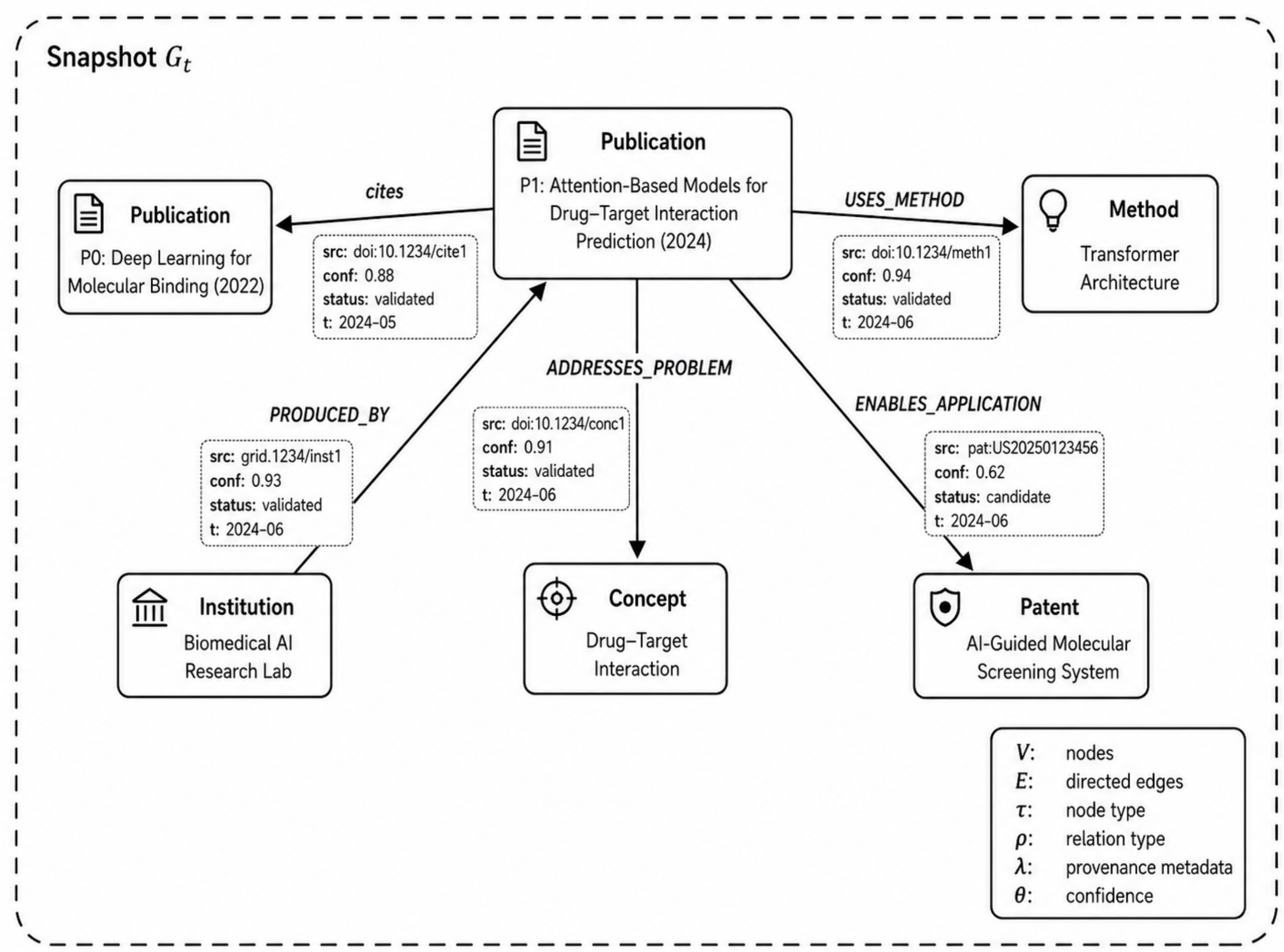

Figure 3 — Dynamic knowledge graph schema with STI example

The graph is *dynamic* in that it is versioned over time. New entities may appear, relations may be added or reinterpreted, and some structures may lose relevance. The framework therefore treats the knowledge graph not as a fixed repository but as a temporally versioned representation of STI activity, enabling queries "as of" any historical state.

At the same time, graph evolution is not assumed to be intrinsically meaningful. Change becomes analytically relevant only when it is recorded through controlled updates that preserve source linkage, timestamping, and validation status. In this way, the dynamic graph remains not only expressive, but reconstructable.

## 4.3 LLM-Assisted Semantic Augmentation

The third layer introduces large language models, but in a deliberately constrained role. Within the framework, LLMs are **not** treated as independent analytical agents or as sources of accepted knowledge. Rather, they function as semantic augmentation instruments that help identify candidate entities, candidate relations, terminological alignments, and possible abstractions from unstructured text.

Their contribution is especially relevant where conventional metadata is insufficient. Abstracts, full texts, patents, grant descriptions, institutional strategies, and policy documents often contain conceptually important relations that remain invisible in standard bibliographic records. LLM-assisted processing makes it possible to surface such relations at a scale that would be difficult to achieve through manual curation or narrowly supervised extraction pipelines alone. However, in the present framework, these outputs remain **provisional**. LLMs generate *candidate enrichments* rather than accepted graph facts. Their role is therefore inferential and assistive, not authoritative. This distinction is central to the architecture. It allows the framework to benefit from the semantic flexibility of LLMs while maintaining a clear separation between generated hypotheses and analytically admissible relations.

Figure 4 presents a parallel swim-lane diagram for all three modules, showing for each the input structure, the prompt constraints applied to the LLM, the form of the structured output, and the handoff arrow to the validation layer. The diagram makes explicit that in all three cases the LLM output is labeled as a candidate enrichment, not an accepted fact, before it reaches the validation stage.

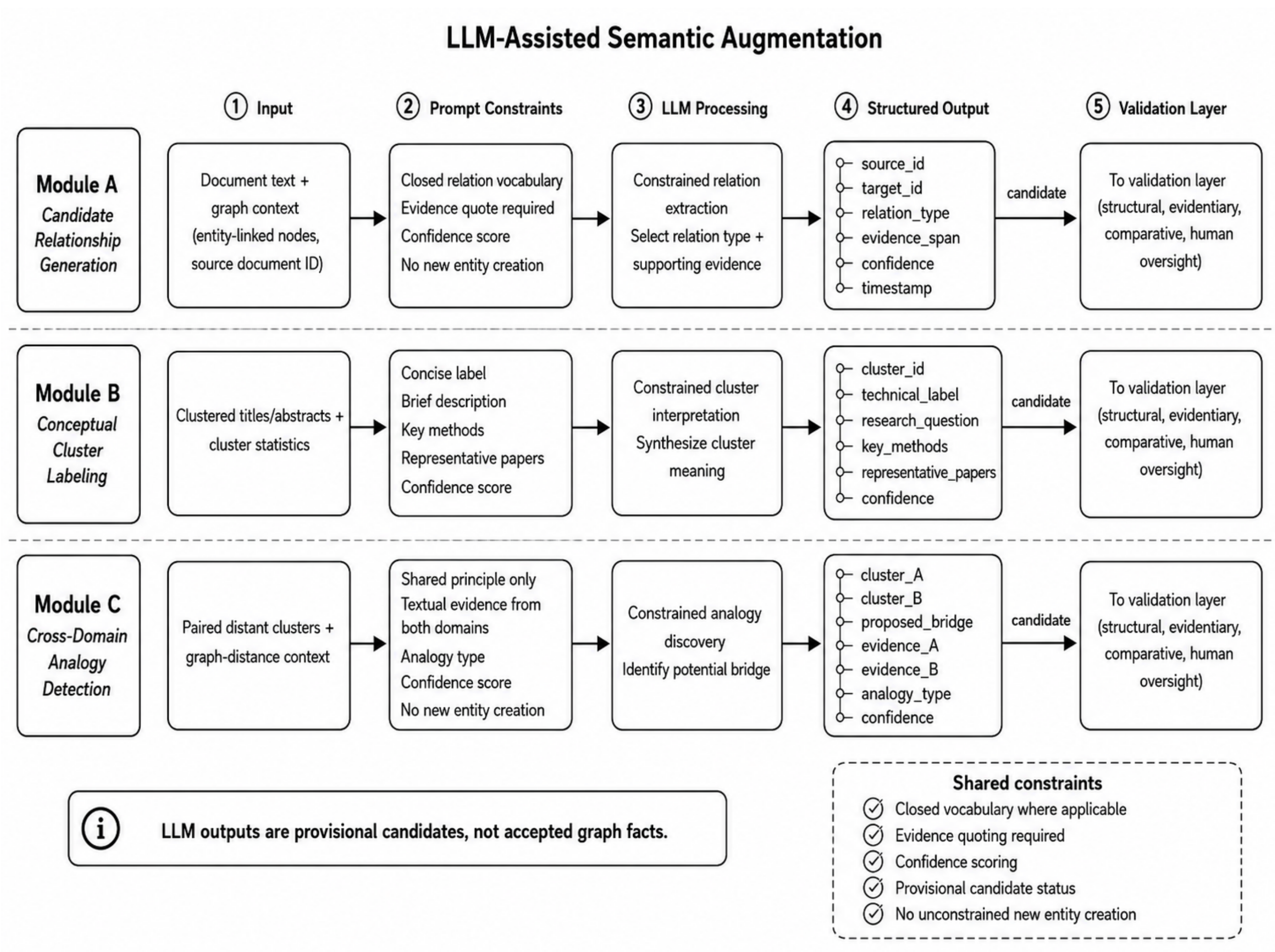

Figure 4 — The three LLM augmentation modules: input–output flow

The framework implements three specific augmentation modules:

**Module A: Candidate Relationship Generation**

This module identifies semantic relationships between existing graph nodes that are implicit in the text but not captured by backbone relations.

*Input:* For each document (e.g., publication abstract, patent claim), the module receives:

- The document text
- The document's node identifier in the graph
- A list of candidate target entities (other nodes) that appear in the same text, obtained through entity linking against the graph

*Prompt design:* The LLM is instructed to choose from a closed set of relation types (e.g., USES_METHOD, ADDRESSES_PROBLEM, CHALLENGES_FINDING, ENABLES_APPLICATION, IMPROVES_UPON, CONTRASTS_WITH), provide a confidence score (0–1), and quote the exact supporting text. "NONE" is a valid output.

*Output structure:*

```
json
{
  "source_id": "pub:10.1234/paper2023",
  "target_id": "concept:graphene_oxide",
  "relation_type": "USES_METHOD",
  "confidence": 0.92,
  "evidence": "We synthesized graphene oxide membranes using modified
Hummers' method...",
  "evidence_span": "sentences 3-4",
  "model": "gpt-4",
  "timestamp": "2024-03-15T10:23:45Z"
}
```

**Module B: Conceptual Cluster Labeling**

This module generates rich semantic descriptions for communities detected in the graph through traditional network analysis (e.g., Louvain or Leiden clustering on co-citation or bibliographic coupling networks).

*Input:* For each detected cluster, the module receives:

- The set of documents (titles and abstracts) belonging to the cluster
- Basic cluster statistics (size, average publication year, dominant venues)

*Prompt design:* The LLM synthesizes the cluster's content into:

- A concise technical label (3–5 words)
- A brief description of the core research question or technological problem
- Key methods, materials, or approaches used
- Three representative papers (DOIs if available)

*Output structure:*

```
json
{
  "cluster_id": "C_1083",
  "technical_label": "Graphene Oxide Water Filtration",
  "research_question": "How can graphene oxide membranes be engineered for
selective ion rejection and high flux in water purification?",
  "key_methods": ["Hummers oxidation", "vacuum filtration", "crosslinking",
"molecular dynamics simulation"],
  "representative_papers": ["doi:10.1021/acs.nanolett.8b01234", ...],
  "confidence": 0.88,
  "model": "gpt-4",
  "timestamp": "2024-03-15T14:17:22Z"
}
```

**Module C: Cross-Domain Analogy Detection**

This module identifies potential connections between distant regions of the knowledge graph, supporting the detection of emerging interdisciplinary links and technological convergence.

Input: Pairs of document clusters from different research areas (identified through graph distance measures or co-citation analysis showing minimal existing connection).

Prompt design: *The LLM is asked to identify shared underlying principles, analogous methodological challenges, or potential technology transfers between the two domains, supported by textual evidence from each.*
Output structure:

```json
{
  "cluster_A": "C_1083",
  "cluster_B": "C_2156",
  "proposed_bridge": "Charge recombination management at interfaces",
  "evidence_A": "Interface engineering with functional groups reduced charge recombination...",
  "evidence_B": "Surface passivation layers minimized non-radiative recombination...",
  "analogy_type": "shared_mechanism",
  "confidence": 0.76,
  "model": "gpt-4",
  "timestamp": "2024-03-16T09:45:12Z"
}
```

Across all modules, constraints ensure LLM outputs remain manageable: closed vocabularies, required evidence, confidence scoring, and prohibition of new entity creation. This modular design also allows future replacement of general-purpose LLMs with smaller, specialized models fine-tuned on validated outputs.

### 4.4 Validation and Governance

The core of the framework is validation. Without it, the architecture would simply reproduce the weaknesses of current LLM-centric workflows. Validation is what transforms semantic possibility into analytically acceptable enrichment. Figure 5 maps this transformation as a sequential funnel through which every candidate triple must pass. Candidates enter at the top (structural validation) and are progressively filtered, flagged, or routed to human expert review before the smallest subset — those that pass all applicable layers — is admitted to the graph with full provenance metadata. The funnel geometry is intentional: it conveys that the validation process is expected to reject a substantial proportion of LLM-generated candidates, and that this attrition is a design feature rather than a failure of extraction.

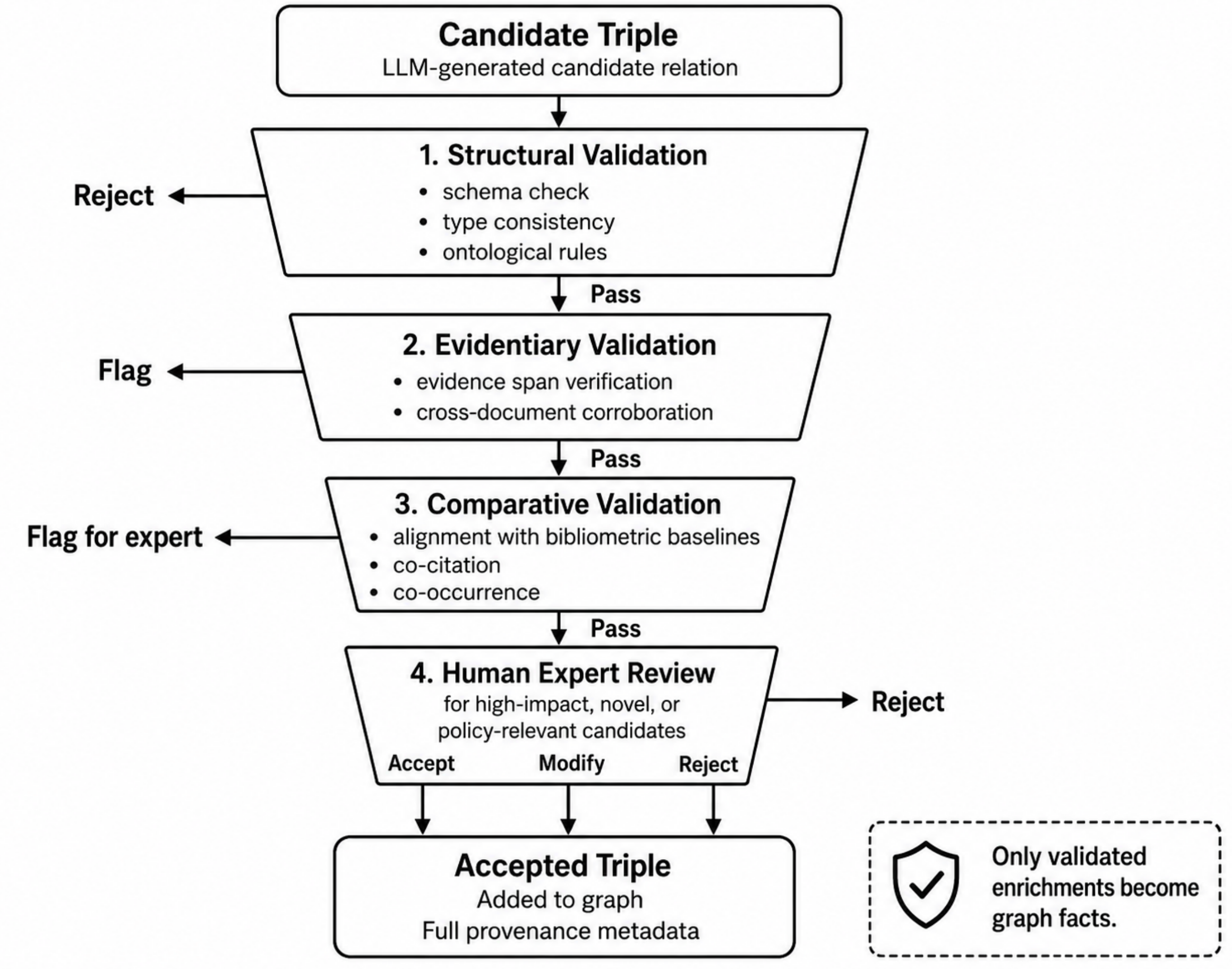

Figure 5 — Multi-Layer validation pipeline

In the proposed framework, validation is **multi-layered**:

1. **Structural validation.** Candidate entities and relations are checked against the graph schema, type ontology, and domain constraints. For example, a relation USES_METHOD is only permitted if the source is a document and the target is a method node. Ontological consistency is enforced using a reasoner or rule-based checks.
2. **Evidentiary validation.** Proposed enrichments are assessed against identifiable source passages. The framework verifies that the quoted evidence actually appears in the source document; it may also require cross-document corroboration (e.g., the same relation proposed in multiple independent texts) and, where available, citation context analysis to see if the relationship is acknowledged elsewhere.
3. **Comparative validation.** Newly introduced graph relations are examined in relation to established bibliometric or network-based signals. For instance, a candidate USES_METHOD edge between a publication and a concept may be compared against co-citation frequency or co-occurrence in authoritative reviews. Candidates that deviate strongly from expected baselines are flagged for additional scrutiny.
4. **Selective human oversight.** Relations with high analytical impact, unusual novelty, low confidence from earlier layers, or potential policy relevance are routed to domain experts. A purpose-built interface presents the candidate, its evidence, and the outputs of automated checks; experts accept, reject, or modify the candidate, and their decisions are recorded as part of the provenance.

Validation decisions are logged, and all accepted enrichments carry metadata indicating which validation layers were passed, the confidence scores, and the final authority (automated or expert). This audit trail enables full inspectability of the graph's construction.

This validation layer also has a governance function. Because STI analytics may inform funding decisions, strategic planning, or policy evaluation, the framework must support accountability beyond technical correctness. Provenance recording, audit trails, confidence marking, and explicit distinction between automatically proposed and validated relations ensure that the process of enrichment itself is inspectable and governable.

## 4.5 Analytical Layer

Once validated, graph enrichments become available for analysis. At this stage, the framework supports both familiar and extended forms of STI analytics.

Familiar analyses include:

- **Citation-based influence** using standard indicators (h-index, field-normalized citation counts)
- **Collaboration structures** (co-authorship networks, institutional partnerships)
- **Institutional productivity** and funding patterns
- **Concept-level network organization** (topic clustering, keyword evolution)

Extended analyses enabled by the enriched graph include:

- **Temporal emergence of research themes** (detection of novel problem–method combinations before citation accumulation)
- **Problem–method trajectories** (tracing how a scientific problem is tackled by different methods over time)
- **Sectoral translation pathways** (tracking the flow of concepts from basic research to patents or industrial applications)
- **Evolution of semantically defined knowledge clusters** (splitting, merging, or shifting of research communities as detected through community detection on the enriched graph)

Crucially, the framework does **not** abandon indicator-based analysis. Rather, it re-situates established indicators within a richer representational environment. Citation structures, publication counts, and collaboration patterns remain useful as baselines and comparators, but they no longer constitute the sole analytical horizon. They coexist with graph-based and semantically enriched analyses that make it possible to ask different questions about STI systems and their evolution.

This design also supports different levels of inference. Some outputs remain **descriptive** (e.g., maps of topic evolution). Others are **diagnostic** (e.g., identifying weak translation between scientific and technological domains). Still others may support **cautious foresight** (e.g., detecting persistent convergence among concepts, sectors, or methods across time). In all cases, the legitimacy of the analysis depends on the traceability of the relations from which it is derived.

## 4.6 Temporal Dynamics and Graph Evolution

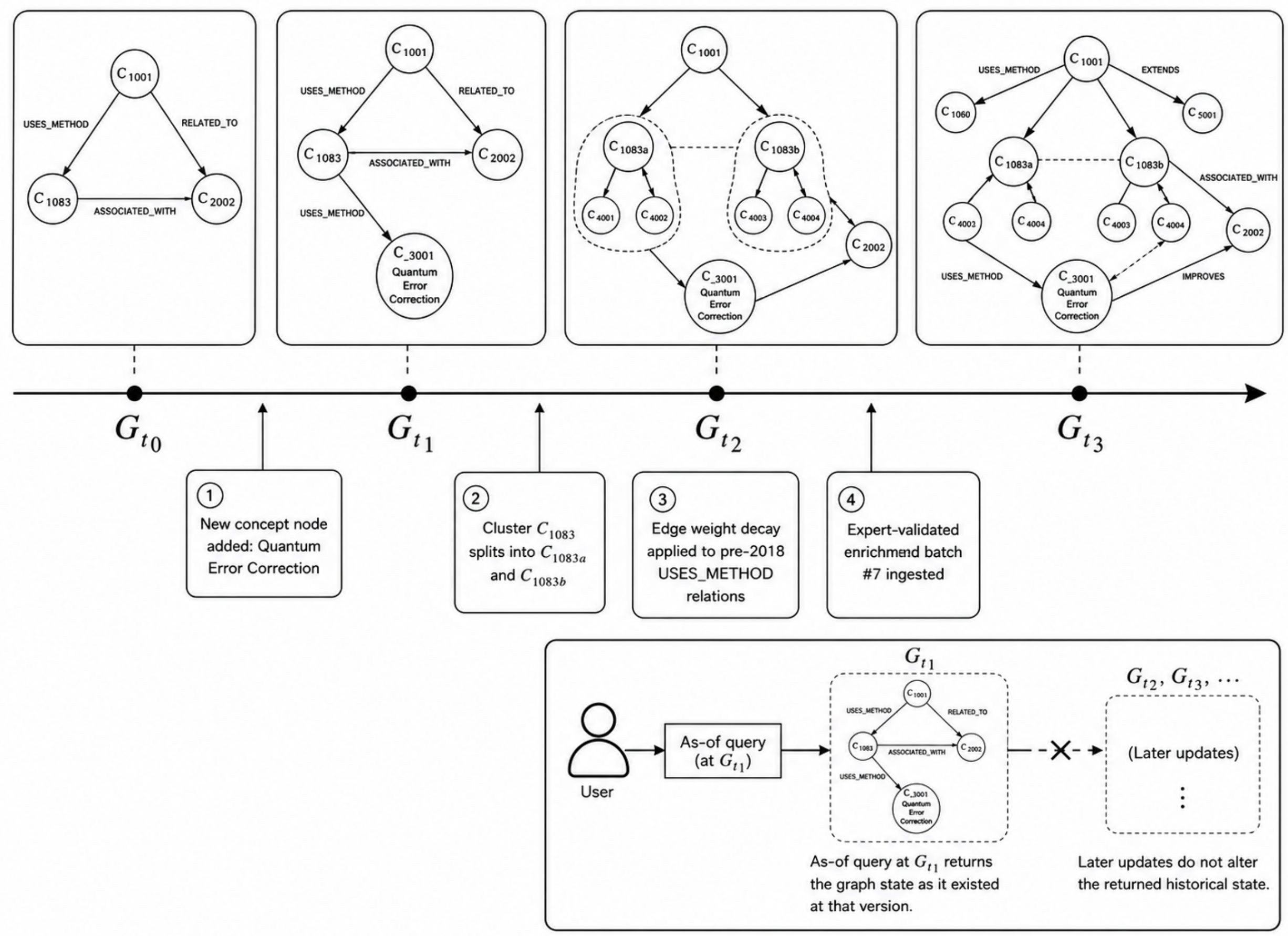

Figure 6 — Graph versioning and temporal evolution

The framework's commitment to temporal accountability requires explicit mechanisms for managing graph evolution over time. Figure 6 illustrates how this commitment is operationalized across a representative update sequence. A horizontal timeline axis shows discrete versioned snapshots $G_{t_0}$ through $G_{t_3}$, with annotated change events between snapshots — including new concept node additions, community splits, expert-validated enrichment batches, and the application of decay functions to older edges. An inset panel demonstrates as-of query semantics: a user querying $G_{t_1}$ receives the graph state as it existed at that version, irrespective of subsequent updates, preserving the reproducibility requirements discussed in Section 3.5.

- **Versioning:** The graph is maintained as a series of versioned snapshots. Each update cycle (e.g., weekly or monthly) produces a new graph version $G_t$. All queries support "as-of" semantics, allowing users to retrieve the graph state at any past version.
- **Provenance tracking:** Every element (node, edge, cluster description) carries metadata about its origin: for backbone relations, source database and extraction timestamp; for validated enrichments, the LLM model, prompt, source documents, validation decisions, and timestamps.
- **Decay and obsolescence:** The framework optionally applies decay functions (e.g., exponential decay with domain-specific half-lives) to edge weights, enabling time-aware ranking that emphasizes recent developments while preserving historical context (Wu et al., 2024).
- **Change detection:** By comparing successive versions, the framework can detect and characterize changes: first appearance of a concept, accretion of relations around an emerging field, reconfiguration of clusters, and semantic velocity (the rate at which a concept migrates from research to application).

### 4.7 Analytical Logic of the Framework

The framework may therefore be summarized as a controlled sequence:

1. **Open scholarly metadata** provide the initial structural substrate.
2. A **dynamic knowledge graph** organizes that substrate into a multi-relational and temporally extensible representation.
3. **LLMs** generate semantically informed candidate enrichments from unstructured sources, constrained by predefined schemas and evidence requirements.
4. **Validation** (structural, evidentiary, comparative, and selective human oversight) determines which candidates are admissible.
5. **Accepted enrichments** are incorporated into versioned graph states with full provenance.
6. **Analytics** are conducted on those graph states, using both traditional indicators and novel graph-based methods.

What distinguishes this sequence from existing approaches is not the presence of each component individually, but their **ordering and epistemic hierarchy**: structure precedes semantic expansion; validation mediates enrichment; analysis follows accepted representation rather than raw generation. This ordering is designed specifically to align semantic scalability with the methodological norms of scientometrics and science-of-science research.

### 4.8 Scope and Intended Use

The framework is not intended as a universal replacement for all STI data systems. Its purpose is narrower and methodological: to provide a disciplined model for integrating semantic augmentation into STI analytics without collapsing the distinction between inference and evidence. It is especially suitable where conventional indicators remain informative but insufficient, and where richer relational representations are needed to understand evolving interactions among research, technology, institutions, and policy.

Equally, the framework should not be understood as fully autonomous. Its design assumes that some degree of oversight remains necessary, particularly in policy-relevant settings. The contribution lies in showing how automation, structure, and validation may be combined in a way that remains compatible with the evidentiary and interpretive standards of science-of-science analysis.

### 4.9 Transition to Application

The framework presented here provides a concrete, operational blueprint for integrating open infrastructures, dynamic knowledge graphs, LLM-based semantic augmentation, and multi-layer validation. The next section illustrates how this logic may support concrete analytical use cases in STI settings, showing both the opportunities and the constraints of semantically enriched and temporally responsive analytics.

## 5. Validation Strategy

A framework that aims to modernize STI analytics must be held to high standards of methodological rigor. Because the proposed hybrid approach combines symbolic knowledge graph structures, temporal modeling, and targeted LLM augmentation, its validation cannot rely on any single metric. Instead, evaluation must demonstrate that the framework produces representations that are simultaneously **semantically richer**, **epistemically more reliable**, and **analytically more useful** than conventional static bibliometrics or ungrounded LLM-driven alternatives.

This section outlines a multi-layered validation strategy designed to assess the framework across four complementary dimensions: extraction validity, structural validity, analytical validity, and operational validity. These layers mirror the architecture of the framework itself and provide a roadmap for future empirical studies. Figure 7 presents these four dimensions as a structured evaluation matrix, with each row corresponding to a validity dimension and columns specifying what is measured, the key metrics, the appropriate comparison baseline, and the primary stakeholder group for whom that dimension matters most. The matrix makes visually apparent that Analytical and Operational validity are the most demanding dimensions — requiring comparison against strong bibliometric baselines and long-term provenance infrastructure, respectively — and are therefore the focus of the staged evaluation agenda presented in Section 5.6.

### 5.1 Extraction Validity

Extraction validity evaluates the quality of LLM-generated candidate entities and relations before any symbolic validation occurs. Since LLMs serve only as semantic augmenters in the framework, the central question is whether they can propose sufficiently accurate candidates to make downstream validation worthwhile.

This dimension should be assessed using expert-annotated gold-standard corpora drawn from heterogeneous STI sources (publication abstracts, full-text articles, patents, and funding announcements). Standard information extraction metrics—precision, recall, and F1-score—should be reported, supplemented by fine-grained error analysis that distinguishes hallucinated relations, incorrect entity typing, unsupported evidence spans, and overly generic outputs.
Domain and genre robustness must also be tested. Performance on scientific papers, patents, and policy documents often differs markedly; the framework must demonstrate acceptable precision across these genres rather than excelling only on clean academic abstracts.

## 5.2 Structural Validity

Structural validity examines whether LLM-generated candidates that pass initial filtering satisfy the formal constraints of the dynamic knowledge graph. This includes ontological type consistency, permitted relation ranges, provenance completeness, temporal ordering constraints, and absence of contradictions in the evolving graph.
Key indicators at this level include:

- Acceptance rate of LLM candidates after symbolic screening;
- Violation rates by relation type and ontological category;
- Graph consistency metrics (e.g., constraint violation frequency, anomalous pattern detection);
- Stability of the graph under incremental updates.

A high rejection rate of implausible candidates would confirm that the symbolic layer meaningfully constrains LLM outputs, while a low rejection rate with preserved analytical utility would indicate effective balance between flexibility and discipline. This layer directly tests the paper's core claim that semantic augmentation must remain subordinate to representational rigor.

## 5.3 Analytical Validity

Analytical validity is the most critical dimension for STI research. It assesses whether the enriched dynamic knowledge graph enables forms of insight that are difficult or impossible to achieve with static bibliometric methods alone.
Evaluation should focus on three representative STI tasks:

- Early detection of emerging interdisciplinary trends and conceptual recombinations;
- Mapping multi-hop science–technology–industry translation pathways;
- Policy-oriented gap analysis and foresight querying.

For each task, the enriched graph should be compared against strong baselines: (1) traditional citation-based indicators from OpenAlex, (2) the static OpenAlex or CS-KG graph without LLM enrichment, and (3) a pure LLM-driven extraction pipeline without symbolic validation. Relevant metrics include lead time in trend detection, accuracy of recombination pathway identification, expert-rated usefulness of policy recommendations, and reduction in hallucinated or unsupported claims.
Success at this level is not defined by marginal gains in precision or recall, but by demonstrable improvements in **interpretive resolution** and **actionable foresight** while preserving auditability.

## 5.4 Operational Validity

Operational validity addresses reproducibility, traceability, and long-term inspectability—essential requirements for policy-relevant STI analytics. The framework must allow users to reconstruct any graph state from preserved source data, model configurations, validation logs, and versioning records.
Key criteria include:

- Full provenance linking every enriched triple back to its source text and validation steps;
- Ability to replay historical graph versions at arbitrary time points;
- Clear distinction between baseline metadata, LLM candidates, validated enrichments, and derived analytics;
- Reproducibility of results across different LLM versions and prompting conditions.

Because STI analytics often inform funding and policy decisions, operational validity should be treated as a core methodological requirement rather than an afterthought.

## 5.5 Expert Assessment and Human-in-the-Loop Evaluation

Automated metrics alone are insufficient for high-stakes domains. Domain experts (scientometricians, STI policy researchers, and subject-matter specialists) should evaluate sampled subgraphs and analytical outputs on factual correctness, substantive meaningfulness, and practical usefulness using structured Likert-scale instruments and qualitative feedback. Hybrid human-LLM disagreement resolution protocols can further quantify the added value of symbolic constraints.

### 5.6 Proposed Staged Evaluation Agenda

Future empirical work should proceed in four progressive stages:

1. **Extraction and structural validity** on annotated benchmark corpora;
2. **Controlled ablation studies** isolating the contribution of symbolic validation versus LLM augmentation;
3. **Analytical validity** through comparative case studies in selected STI domains (e.g., AI for sustainability, quantum technologies);
4. **Longitudinal operational validity** assessing reproducibility and auditability under realistic data-update scenarios.

This staged approach allows precise identification of failure points and ensures that improvements at lower layers translate into genuine gains at the analytical and operational levels.

The validation strategy outlined here reinforces the central thesis of the paper: meaningful modernization of STI analytics requires more than powerful language models or larger graphs. It demands a disciplined integration of semantic power with symbolic structure, temporal awareness, and rigorous accountability mechanisms. By specifying concrete criteria and a clear evaluation roadmap, this section provides both a methodological foundation for the proposed framework and a benchmark against which future implementations can be judged.

| Validity Dimension (Row) | What is measured | Key metrics (examples) | Comparison baseline | Primary stakeholder |
|---|---|---|---|---|
| **1. Extraction Validity** Accuracy of LLM-extracted entities and relations relative to source texts. | • Correctness of extracted entities and relations<br>• Evidence grounding<br>• Absence of hallucinations<br>• Extraction completeness | • Precision / Recall / F1<br>• Evidence span accuracy (%)<br>• Hallucination rate (%)<br>• Type assignment accuracy (%)<br>• Coverage of gold-standard set | • Human-annotated gold standard<br>• Existing NER/RE systems<br>• Manual curation samples | Data curators, Ontology engineers, ML/NLP specialists |
| **2. Structural Validity** Consistency of the graph with schema, ontology and temporal rules. | • Schema conformity<br>• Type and domain/range consistency<br>• Temporal consistency<br>• Provenance completeness<br>• Constraint satisfaction | • Schema violation rate (%)<br>• Ontology rule violation rate (%)<br>• Temporal inconsistency rate (%)<br>• Provenance completeness (%)<br>• Constraint satisfaction score | • Ontology/Schema constraints<br>• Knowledge graph best practices<br>• Prior graph versions | KB engineers, Ontology managers, Data stewards |
| **3. Analytical Validity** Extent to which the graph enables meaningful insight and improves analytical outcomes. | • Added interpretive resolution<br>• Early detection of emerging themes<br>• Discovery of cross-domain linkages<br>• Science–technology–policy pathway usefulness | • Time-to-detection gain (months)<br>• Increase in unique pathways (%)<br>• Cluster/topic coherence (e.g., NMI, modularity)<br>• Novel linkage yield (validated)<br>• Expert usefulness rating | • Traditional bibliometric indicators<br>• Static citation networks<br>• Baseline KG without LLM augmentation | Researchers, Analysts, Policy analysts |
| **4. Operational Validity** Reliability, reproducibility, and sustainability of the framework in practice. | • Reproducibility of results<br>• Temporal "as-of" correctness<br>• System stability under updates<br>• Traceability and auditability<br>• Scalability and performance | • Reproducibility rate (%)<br>• As-of query accuracy (%)<br>• Update stability score<br>• Provenance traceability (%)<br>• Throughput / latency metrics | • Re-execution on same data<br>• Prior versions (as-of comparisons)<br>• Alternate pipelines / models<br>• System performance targets | System operators, Platform managers, Auditors |

| Relative difficulty to evaluate* | Extraction Validity | Structural Validity | Analytical Validity | Operational Validity | |
|---|---|---|---|---|---|
| | Relatively easier (Direct text grounding and annotation possible) | Moderate (Deterministic rules and schema checks) | Challenging (Dependent on use cases and expert judgment) | Most challenging (System-level, longitudinal, and context-dependent) | *Difficulty reflects typical evaluation complexity and subjectivity. |

Figure 7 — Validation strategy evaluation matrix

## 6. Discussion, limitations, and implications

The framework proposed in this paper is intended as a methodological extension of STI analytics rather than as a replacement for existing bibliometric practice. Its central claim is that semantically enriched, temporally responsive, and structurally validated knowledge representations can improve the analytical reach of STI research without abandoning the methodological discipline that gives scientometric approaches their credibility. At the same time, the framework introduces new complexities, both technical and epistemic, that require careful interpretation. This section discusses the significance of the framework, its limitations, its governance implications, and the conditions under which it may be most appropriately used.

## 6.1 From indicator-based evaluation to representationally enriched analytics

A central implication of the framework is that the modernization of STI analytics should not be understood simply as the adoption of more advanced AI tools. Rather, it should be understood as a shift in representational logic. Traditional bibliometric indicators remain valuable because they provide standardized, reproducible, and institutionally legible measures of scientific activity. Their limitation is not that they are invalid, but that they operate on abstractions that are often too coarse for understanding the semantic and temporal dynamics of contemporary STI systems.

The proposed framework suggests that these limitations can be addressed not by discarding indicator-based traditions, but by situating them within a broader analytical environment. In this environment, bibliometric indicators become one layer of evidence among others: not the sole output of STI analytics, but one validated baseline within a richer representational structure. This repositioning is methodologically important because it allows the field to expand its analytical scope without dissolving its existing standards of evidence.

In that sense, the framework is best understood as a move from indicator-centered STI evaluation toward representationally enriched STI analysis. The emphasis shifts from counting and ranking alone to the structured interpretation of evolving relations among concepts, institutions, methods, technologies, and policy priorities. Figure 8 renders this shift as a two-panel before/after diagram. The left panel represents traditional STI analytics: a flat layer of bibliometric indicators — citation counts, h-index, co-authorship ties — derived directly from publication metadata, with their characteristic weaknesses of temporal lag and semantic shallowness annotated. The right panel shows the proposed framework, in which those same bibliometric indicators are preserved as a validated baseline layer, now situated within a richer representational environment that additionally supports typed semantic relations, temporal versioning, and policy-oriented gap analysis. A central axis labeled 'Representational Extension, Not Replacement' connects the two panels, making visually explicit the paper's core methodological commitment: this is an augmentation of bibliometric reasoning, not a repudiation of it.

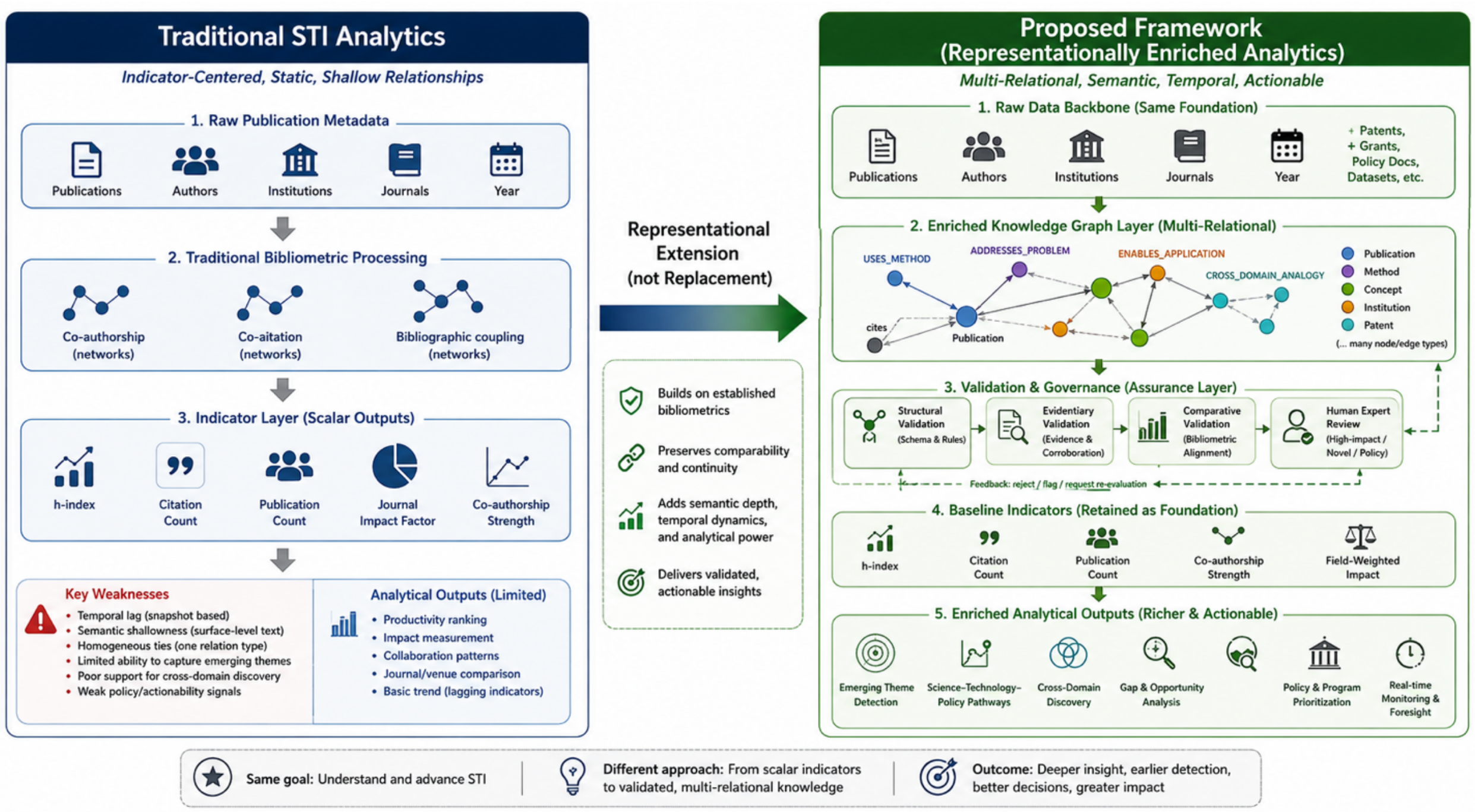

Figure 8 — Conceptual Shift: From Indicator-Centered to Representationally Enriched Analytics

## 6.2 The value and limits of LLM-assisted semantic augmentation

The framework treats LLMs as enabling technologies for semantic augmentation, not as autonomous analytical authorities. This distinction is fundamental. The appeal of LLMs lies in their ability to process heterogeneous and unstructured textual material at a scale that would be difficult to achieve through manual curation or narrowly supervised extraction alone. In the STI context, this creates the possibility of representing implicit conceptual relations, cross-domain linkages, and policy-relevant alignments that remain largely invisible in conventional metadata.

However, the value of LLM-based augmentation is inseparable from its limitations. Probabilistic language models do not produce stable or inherently verifiable representations. Their outputs may vary across prompts, model versions, and domains; they may overgeneralize, hallucinate, or reproduce asymmetries present in the underlying corpus. For this reason, the framework does not treat semantic generation as evidence in itself. Its usefulness depends entirely on the quality of the validation process through which candidate enrichments are filtered, contextualized, and made analytically admissible.

This has an important implication for STI analytics. The practical value of LLMs does not lie in replacing scientometric judgment with automated reasoning, but in widening the space of candidate relations that can then be examined under structured constraint. The framework therefore aligns with a broader methodological principle: AI should increase analytical reach without displacing evidentiary responsibility.

## 6.3 Why validation is not enough on its own

Although Section 5 established validation as the mediating principle of the framework, validation alone does not resolve all methodological concerns. A relation may pass structural and evidentiary checks and still contribute little to substantive interpretation. Likewise, a graph may be internally consistent while remaining analytically uninformative or socially biased. In other words, admissibility is not equivalent to usefulness.

This matters because the proposed framework could otherwise be misunderstood as a technical solution to an essentially interpretive problem. STI analytics is not only about extracting more relations from more data; it is also about determining which relations matter, in which contexts, and for which decisions. Validation can determine whether an enrichment is acceptable within the framework, but it cannot by itself determine whether the resulting analysis is meaningful, proportionate, or appropriate for policy use.

For this reason, the contribution of the framework should be interpreted carefully. It offers a disciplined way to construct richer analytical representations, but it does not eliminate the need for domain expertise, interpretive caution, or methodological pluralism.

## 6.4 Reproducibility, temporality, and the burden of dynamic systems

One of the central promises of the framework is that it supports temporally responsive STI analytics. Yet this promise comes with a cost. Dynamic and versioned graph systems are inherently more difficult to reproduce and maintain than static bibliometric snapshots. As graph states evolve, model configurations change, and source infrastructures are updated, the burden of methodological accountability increases rather than decreases.

This is especially relevant in contexts where analytical outputs may be revisited after the fact, such as funding evaluations, strategic planning exercises, or policy audits. In such settings, it is not enough for the framework to produce plausible outputs in the present. It must also preserve a sufficiently complete record of how those outputs were generated, which graph state they depended on, which source materials were used, and which validation decisions were applied. Without this, the move to dynamic knowledge representations risks undermining one of the core strengths of conventional scientometrics: reproducible interpretability.

The framework addresses this problem through versioning, provenance, and explicit separation between generated candidates and accepted relations. Even so, these mechanisms increase the operational demands of the system. A dynamic STI analytics environment may be more analytically powerful, but it is also more demanding in terms of curation, governance, and methodological discipline.

## 6.5 Bias, visibility, and uneven knowledge infrastructures

A further limitation concerns the underlying data environment. No STI analytics framework operates in a neutral informational space. Open scholarly infrastructures, patent records, grant datasets, and policy documents all reflect uneven distributions of visibility across regions, languages, disciplines, institutions, and publication traditions. LLMs may amplify these asymmetries by producing better semantic outputs for well-represented domains and weaker outputs for marginal or under-documented ones.

This has both analytical and normative consequences. Analytically, it means that semantically enriched graphs may be more complete for some domains than for others. Normatively, it means that decisions informed by such systems may reproduce existing inequalities in what counts as visible, connected, or strategically important. A framework that appears more comprehensive than traditional bibliometrics may, in practice, remain constrained by the same structural absences—while making them harder to detect because of its richer surface representation.

Accordingly, applications of the framework should include explicit acknowledgment of corpus boundaries, language limitations, coverage gaps, and known infrastructural asymmetries. If the framework is to support evidence-based STI policy, it must also support transparency about the evidentiary conditions under which its insights are produced.

## 6.6 Appropriate use cases and boundary conditions

The framework is not equally appropriate for all analytical contexts. It is most suitable where traditional indicators remain informative but insufficient—particularly where the analytical goal involves understanding conceptual evolution, science–technology linkages, emerging interdisciplinary recombinations, or the alignment between research developments and policy priorities. In such settings, semantically enriched graph representations may offer clear advantages over citation-based summaries alone.

By contrast, the framework is less clearly advantageous in settings where highly standardized and institutionally stabilized indicators are required, where source data are sparse or highly uneven, or where the costs of interpretive error are such that only conservative representations are acceptable. It is also not intended as a generic replacement for domain-specific ontology engineering or highly optimized information extraction systems. Its purpose is broader and methodological: to support STI analytics under conditions where semantic richness, temporal responsiveness, and evidentiary control must be balanced.

Recognizing these boundary conditions is important because it prevents the framework from being interpreted as a universal solution. Its contribution lies in clarifying one viable path for extending STI analytics, not in claiming that all such analytics should now be reorganized around LLM-assisted knowledge graphs.

## 6.7 Governance implications for STI analytics

Because STI analytics often informs funding, strategy, and policy, governance must be treated as intrinsic to the methodology rather than as a downstream implementation concern. In a framework of this kind, governance is not only about access control or responsible AI in the abstract; it is about preserving distinctions that are methodologically essential. Users must be able to distinguish among baseline metadata, machine-generated candidates, validated enrichments, and graph-derived interpretations. Confidence markings, provenance records, validation histories, and version identifiers should therefore be treated as part of the analytical object, not as auxiliary metadata.

There is also a broader governance implication. As STI analytics becomes more semantically expressive, the risk increases that generated structure will be mistaken for established evidence. In policy settings, this distinction is particularly important. A semantically plausible relation may be valuable as a hypothesis or exploratory signal, yet still be inappropriate as a basis for formal decision-making. The framework proposed here attempts to preserve that distinction through validation and traceability, but institutional users must still exercise caution in how graph-derived outputs are interpreted and used.

In this respect, the framework points toward a wider methodological need in STI analytics: the development of governance standards for semantically enriched, AI-assisted analytical systems. Such standards would need to address not only technical accuracy, but also traceability, interpretability, bias disclosure, and the appropriate use of machine-assisted inference in evaluative contexts.

## 6.8 Implications for future research

The framework also implies a broader research agenda. One direction concerns empirical testing: the validation strategy outlined in Section 5 should be implemented across multiple STI domains and compared systematically with conventional bibliometric approaches. A second concerns methodological benchmarking, particularly for relation types that are likely to be unstable, weakly observable, or domain-sensitive. A third concerns interface and interpretability research, especially how analysts and policymakers interact with graph-based and semantically enriched STI representations.

A further direction concerns evaluation theory. If STI analytics is to move toward richer representations, then its evaluative concepts may also need refinement. Established ideas such as interdisciplinarity, emergence, translation, and policy alignment may require new operationalizations when analyzed through dynamic and semantically enriched graph structures. The modernization of STI analytics is therefore not only a matter of new data infrastructures and AI tools; it may also reshape the conceptual vocabulary through which STI systems are interpreted.

## 6.9 Concluding perspective

The framework proposed in this paper should therefore be understood as a disciplined but partial step toward modernizing STI analytics. Its value lies in showing how open scholarly infrastructures, dynamic knowledge graphs, constrained LLM-based semantic augmentation, and multi-layer validation may be integrated within a common methodological logic. Its limitations lie in the fact that each of these components also introduces new burdens: of validation, governance, reproducibility, and interpretive care.
The significance of the framework is not that it resolves the tension between semantic richness and methodological control once and for all. Rather, it provides a structured way of managing that tension. In doing so, it offers a plausible foundation for next-generation STI analytics that is more semantically expressive and temporally responsive than conventional bibliometrics, while remaining aligned with the evidentiary norms of science-of-science research.

# 7. Conclusion

This paper has argued that the modernization of science, technology, and innovation (STI) analytics requires more than incremental refinement of existing bibliometric indicators. While traditional bibliometric and scientometric approaches remain indispensable for standardized, reproducible, and institutionally legible evaluation, they are limited in their ability to represent the semantic and temporal complexity of contemporary STI systems. In response to this limitation, the paper proposed a framework that integrates three complementary components: bibliometric baselines as validated indicators, dynamic knowledge graphs as relational and temporally extensible representational structures, and large language models as constrained instruments of semantic augmentation.
The central contribution of the paper is therefore neither a rejection of bibliometrics nor an endorsement of unconstrained AI-driven analytics. Rather, it is the articulation of a principled middle path. In this formulation, LLMs are positioned not as autonomous analytical authorities, but as assistive mechanisms for surfacing candidate semantic relations from heterogeneous and unstructured STI-relevant data sources. Knowledge graphs provide the structural substrate through which these relations can be represented, versioned, and integrated. Validation functions as the mediating principle that determines which enrichments become analytically admissible. Taken together, these elements define an STI analytics architecture that aims to be semantically richer, temporally more responsive, and methodologically more accountable than either static bibliometrics or ungrounded LLM-based workflows alone.
A second contribution of the paper is methodological. By explicitly distinguishing among theoretical foundations, framework architecture, and validation strategy, the paper clarifies the conditions under which semantically enriched STI analytics might become credible within science-of-science research. In particular, it argues that the value of such a framework cannot be assessed through extraction accuracy alone. Any serious evaluation must also address structural validity, analytical usefulness, and operational accountability, including reproducibility, provenance, and the inspectability of graph states over time. This position is especially important in policy-relevant settings, where STI analytics may inform funding decisions, strategic planning, or institutional assessment.
The broader implication is that the future of STI analytics may depend less on choosing between legacy indicators and advanced AI than on developing frameworks capable of integrating both under explicit methodological constraint. If successful, such integration would allow STI systems to be analyzed not only as collections of outputs and impacts, but as evolving relational environments in which concepts, institutions, technologies, and policy priorities interact over time. At the same time, the paper has emphasized that greater semantic and temporal richness also increases the burden of validation, governance, and interpretive caution. For that reason, the proposed framework should be understood as a disciplined extension of existing analytical traditions, not as a substitute for them.
The paper has remained intentionally conceptual. It does not claim to provide a fully implemented or empirically validated system. Instead, it offers a theoretically grounded and methodologically structured foundation on which future empirical work can proceed. The next steps are therefore clear: to test the framework across heterogeneous STI domains, to benchmark its outputs against conventional bibliometric approaches, and to assess whether its enriched representations genuinely support better and more accountable forms of STI analysis.
In this sense, the paper's main claim is ultimately modest but important. Contemporary STI analytics requires richer representations than static indicators alone can provide, but such enrichment becomes legitimate only when it remains subordinate to validation, provenance, and methodological accountability. The framework proposed here is one attempt to meet that requirement. By bringing together bibliometric reasoning, dynamic knowledge graph modeling, and constrained LLM-based semantic augmentation, it contributes a possible foundation for a next generation of STI analytics that is both more expressive and more disciplined.

## Statements and Declarations

**Competing Interests:** The authors declare that they have no financial or non-financial interests, direct or indirect, that are related to the work submitted in this manuscript. The authors have no relevant financial interests, memberships, affiliations, employment, consultancies, stock ownership, honoraria, grants, or expert testimony to disclose in relation to the content of this article.

**Funding:** This research received no specific grant from any funding agency in the public, commercial, or not-for-profit sectors.